\documentclass[aps,amsmath,twocolumn,amssymb,floatfix,showpacs,superscriptaddress,nofootinbib,longbibliography]{revtex4-1}
\usepackage[dvips]{graphics}
\usepackage{bm}
\usepackage{MnSymbol}
\usepackage{braket}
\usepackage[dvipsnames]{xcolor}
\usepackage{float}
\usepackage{subfigure}
\usepackage{tikz}
\usepackage{epsfig,amssymb,amsmath,latexsym}
\usepackage[colorlinks=true,linktoc=page,linkcolor=magenta,citecolor=blue,urlcolor=purple]{hyperref}

\mathchardef\mhyphen="2D 

\newcommand{\ie}{{ i.e.,\,\,}}
\newcommand{\eg}{{e.g.,~}}

\newcommand{\la}{{\langle}}
\newcommand{\ra}{{\rangle}}

\newcommand{\bea}{\begin{eqnarray}}
\newcommand{\eea}{\end{eqnarray}}
\newcommand{\beq}{\begin{equation}}  
\newcommand{\eeq}{\end{equation}}

\newcommand{\non}{\nonumber}  
\newcommand{\mc}{\mathcal}
\newcommand{\mbf}{\mathbf}

\newcommand{\up}{\uparrow}
\newcommand{\dn}{\downarrow}

\usepackage[normalem]{ulem}
\definecolor{lime}{HTML}{A6CE39}
\usepackage{sidecap,tikz}
\DeclareRobustCommand{\orcidicon}{\hspace{-1.0mm}
	\begin{tikzpicture}
		\draw[lime, fill=lime] (0.0,0.0) 
		circle [radius=0.15] 
		node[white] {{\fontfamily{qag}\selectfont \tiny \,ID}};
		\draw[white, fill=white] (-0.0525,0.095) 
		circle [radius=0.007];
	\end{tikzpicture}
	\hspace{-3.0mm}
}
\foreach \x in {A, ..., Z}{\expandafter\xdef\csname orcid\x\endcsname{\noexpand\href{https://orcid.org/\csname orcidauthor\x\endcsname}{\noexpand\orcidicon}}
}

\AtBeginDocument{%
	\newwrite\bibnotes
	\def\bibnotesext{Notes.bib}
	\immediate\openout\bibnotes=\jobname\bibnotesext
	\immediate\write\bibnotes{@CONTROL{REVTEX41Control}}
	\immediate\write\bibnotes{@CONTROL{%
			apsrev41Control,author="08",editor="1",pages="1",title="1",year="1"}}
	\if@filesw
	\immediate\write\@auxout{\string\citation{apsrev41Control}}%
	\fi
}%

\begin{document}

\title{Fermi arcs mediated transport in inversion symmetry-broken Weyl semimetal nanowire and its hybrid junctions}  

\author{Amartya Pal\orcidA{}}
\email{amartya.pal@iopb.res.in}
\affiliation{Institute of Physics, Sachivalaya Marg, Bhubaneswar-751005, India}
\affiliation{Homi Bhabha National Institute, Training School Complex, Anushakti Nagar, Mumbai 400094, India}

\author{Paramita Dutta\orcidB{}}
\email{paramita@prl.res.in}
\affiliation{Theoretical Physics Division, Physical Research Laboratory, Navrangpura, Ahmedabad-380009, India}

\author{Arijit Saha\orcidC{}}
\email{arijit@iopb.res.in}
\affiliation{Institute of Physics, Sachivalaya Marg, Bhubaneswar-751005, India}
\affiliation{Homi Bhabha National Institute, Training School Complex, Anushakti Nagar, Mumbai 400094, India}

\begin{abstract}
The emergence of gapless surface states, known as Fermi arcs (FAs), is one of the unique properties of the novel topological Weyl semimetal (WSM). However, extracting the signatures of FAs from the bulk states has always been a challenge as both of them are gapless in nature and connected to each other. We capture the signatures of FAs via transport in an inversion symmetry (IS)-broken WSM. We study the band structure and the properties of FAs like shape, spin polarization considering slab and nanowire (NW) geometry, and then compute the two-terminal conductance in WSM NW in terms of the scattering coefficients within the Landauer formalism. We find the FA-mediated conductance to be quantized in units of $2e^2/h$. We extend our study to the transport in WSM/Weyl superconductor (WSC) NW hybrid junction using the Blonder-Tinkham-Klapwijk (BTK) formalism. We show that due to the intricate spin textures, the signatures of the FAs can be captured via Andreev reflection process. We also show that our results of conductance are robust against delta-correlated quenched disorder and thus enhancing the experimental feasibility.
\end{abstract}

\maketitle

\section{Introduction}
Over the past decade, Weyl semimetal (WSM) has emerged as a novel gapless topological phase in three dimensional ($3$D) semimetals due to its nontrivial band structure and intriguing transport properties~\cite{Armitage2018,Rao2016,McCormick2017,Yan2017,Burkov2018}. Weyl fermions were initially proposed as a solution of massless Dirac equation in 1929~\cite{Dirac1928,Weyl1929}. Later, these fermions as low energy excitations, have been theoretically proposed in $3$D topological insulators at the transition phase between the trivial and nontrivial insulating phases by breaking either time reversal symmetry (TRS) or inversion symmetry (IS)\,\cite{Murakami2007,Vazifeh2013}, topological insulator heterostructures\,\cite{Burkov2011a,Burkov2011b,Zyuzin2012a,Zyuzin2012b,Hal2012,Liu2013}, pyrochlore iridates~\cite{Wan2011,Bzdu2015,Witczak2012} etc. The experimental realizations of WSM phase in real materials \eg TaAs, TaP, NbAs, NbP, WTe$_2$, magnetic Heusler materials etc.~\cite{Lv2015a,Lv2015b,Lv2015c,Lu2015,Xu2015a,Xu2015b,Xu2015c,Soluyanov2015,Moll2016,Wang2016,Xu_2015,Ojanen2013,Sun2015,Feng2016,Xu2016,Potter2014,Xu_2015,Xu2016} have opened up the opportunity for the plethora of research works for both theorists and experimentalists.

WSMs exhibit a unique bulk band structure where the valence and conduction bands intersect at an even number (minimum two for TRS-broken and four for IS-broken WSM) of isolated points in momentum space known as Weyl nodes (WNs). Around these WNs, the bulk bands disperse linearly with momentum, resembling the dispersion of $3$D massless relativistic fermions. WNs are recognized as the monopoles of Berry curvature in momentum space, while their charge, termed chirality, determines their topological nature~\cite{Armitage2018,Rao2016,McCormick2017,Yan2017,Burkov2018}. Importantly, WNs always apear pairwise with opposite chiralities, ensuring a net zero chiral charge over the entire Brillouin zone. Effects of both weak and strong disorder~\cite{Altland2016,Shapourian2016,Klier2019,Sbierski2014,Chen2015}, and interactions~\cite{Maciejko2014,Witczak2014,Hosseini2015,Jacobs2016,Laubach2016,Roy2017,Boettcher2020} on the WSM phase have been investigated.

In addition to the nontrivial bulk bands, an intriguing and exotic aspect of WSMs is the presence of nontrivial surface states, known as Fermi arcs (FAs)~\cite{Lv2015a,Lv2015b,Lv2015c,Moll2016,Xu2015c,Xu2015a,Xu2015b,Xu2011,Sun2015a,Sun2015b,Wang2018,Chen2020,Li2015}. When projected onto a surface Brillouin zone (sBZ), these surface states manifest as arcs with their endpoints located at the projection of the bulk WNs on the sBZ. Close to the projection of the WNs, FAs states can leak into the bulk and reappear at the opposite sBZ~\cite{Armitage2018,Rao2016,McCormick2017}. Notably, both surface and bulk states of WSMs are gapless. This is in sharp contrast to the $3$D topological insulators, where gapless surface states lie within the bulk gap and are exponentially localized near the surface. Signatures of these FAs in transport properties have been investigated both theoretically~\cite{Baireuther2016,Gorbar2016,Igarashi2017,Breitkreiz2019,Mukherjee2019,Kaladzhyan2019} and experimentally~\cite{Li2015,Wang2016,Wang2018,Chen2020}.

Due to the gaplessness nature, separating these surface states from the bulk and identifying the sole signatures of FAs in the  transport measurement has always been a challenging task. Very recently, it has been shown that bulk states are gapped out in TRS-broken WSM nanowire (NW) due to its finite size effect. Within the bulk confinement gap, only surface states are present and their contributions to the conductance becomes quantized in units of $e^2/h$ in a two-terminal setup~\cite{Kaladzhyan2019}. However, most of the WSM phases in reality are observed to be ISB because of the plenty of crystal structure asymmetries found in nature.


In recent years, investigation of transport properties in superconducting hybrid junctions of WSMs have attracted significant attention due to the interplay of superconductivity and the non-trivial topology of WSMs~\cite{Meng2012,Cho2012,Uchida2014,Khanna2016,Mukherjee2017,Zhang2018a,Zhang2018b,
Saxena2023,Dutta2020,Saxena2023,Chatterjee2023}. Most of these studies concern the bulk properties of WSM. Transport phenomena become much more subtle and fascinating when the role of surface states are also taken into account~\cite{Khanna2014,Bednik2015,Chen2016,Baireuther2017, Dutta2019,Faraei2019,Zheng2021}. Till date, there are a few studies on FA mediated transport phenomena exist in the  literature~\cite{Baum2015,Kaladzhyan2019,Pareek2018,Kumari2024,Kuibarov2024}. Particularly, signatures of FAs in IS broken (ISB) Weyl NW hybrid junctions have not been investigated so far in the literature, to the best of our knowledge. Since NW junctions are very useful in separating the contributions due to FAs, it remains interesting to look for the role of FAs in transport phenomena via the 
Andreev reflection (AR) in Weyl NW-based superconducting hybrid junctions.

\begin{figure}
	\includegraphics[scale=0.3,page=1]{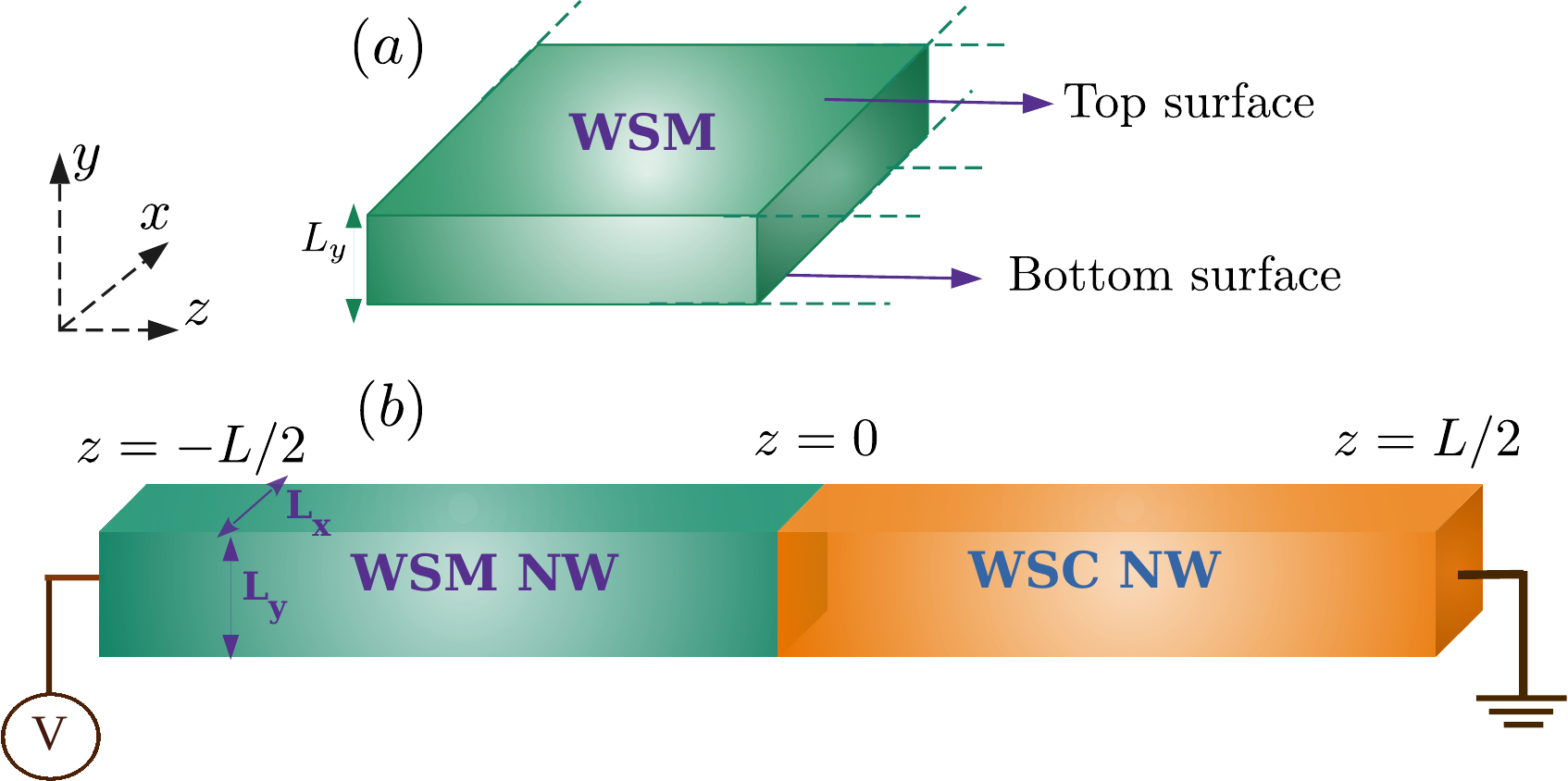}
	\caption{(a) Schematic diagram of a WSM slab geometry with translational invariances along 
          $x$- and $z$-directions, but a finite width ($L_y$) along $y$-direction.  
	(b) Schematic diagram of a WSM NW ($-L/2<z<0$)/WSC NW ($L/2>z>0$) hybrid junction in the presence of a voltage bias $V$, where $L_x$ and $L_y$ denote the finite sizes along $x$ and $y$-directions, respectively.}
	\label{Fig:Schematic}
\end{figure}

With these motivations, in this present article, we investigate ISB WSM NW in a two-terminal set-up 
in two conditions: (i) bare NW and (ii) its hybrid junction with superconducting pairing (WSC) tailoring a WSM/WSC NW junction as shown in Fig.~\ref{Fig:Schematic}. Here, superconductivity in WSC NW 
can be generated either via the proximity effect~\cite{Khanna2014} or electron-electron correlation~\cite{Bednik2015,Sekine2013,Qin2019,Cho2012}.
In our work, we address the following intriguing questions: $(1)$ Is it possible to separate out the contributions of FAs from the bulk states in ISB WSM NW, similar to the TRS-broken WSM case? 
$(2)$ Is it possible to capture the signature of FAs via the Andreev process in such hybrid junction? 
$(3)$ Does the transport signature become quantized in ISB WSM junctions too? $(4)$ Are these quantizations robust against disorder? 

The rest of the article is organized as follows. In Sec.\,\ref{Sec:II}, we introduce our model, compute and analyse the band structure and FAs in slab and NW geometry. In Sec.\,\ref{Sec:III}, we investigate the conductance in WSM NW and WSM/WSC NW hybrid juction, and subsequently in Sec.\,\ref{Sec:IV}, we check the robustness of our results against the random onsite disorder. Finally, in Sec.\,\ref{Sec:V}, we summarize and conclude our results.

\section{Model and Band Structure}\label{Sec:II}
In this section, we first introduce our model of ISB WSM and WSM with 
superconducting correlation. For the discussion on the WSM phase in detail, we show the bulk band structure, followed by a thorough discussions on the surface states. To obtain the surface states, we require finite boundary which can be achieved by making the WSM finite atleast along one direction. 
\begin{figure}
	\includegraphics[scale=0.35,page=1]{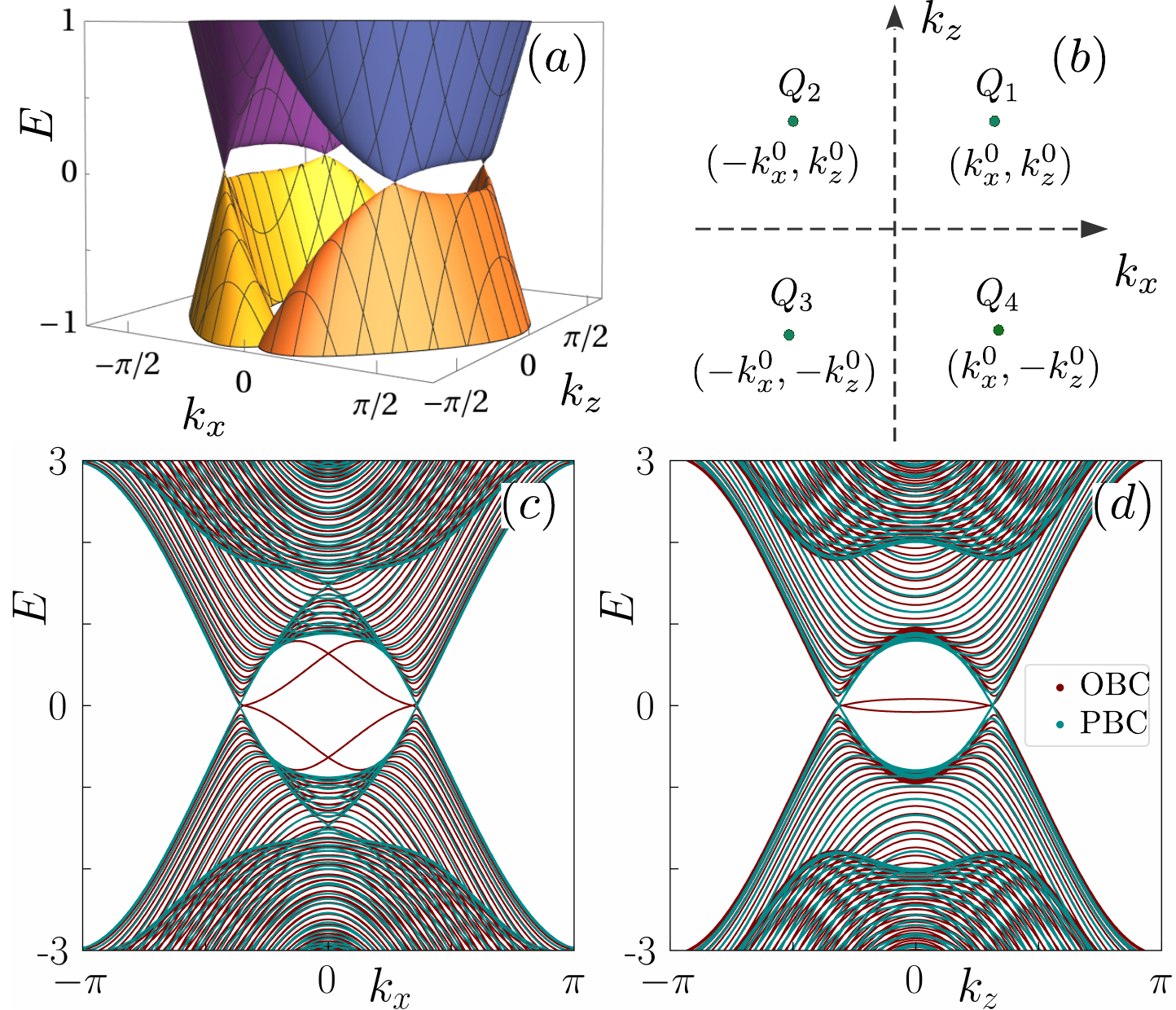}
	\caption{(a) Bulk band structure of the ISB WSM is depicted in the $k_y=0$ plane. (b) Locations 
	of the four WNs are shown in the $k_x-k_z$ plane choosing $k_y=0$. (c-d) Band structure of WSM considering slab geometry is demonstrated choosing $L_y= 40$ 
	and considering both OBC and PBC along $y$-direction.}
	\label{Fig:Bulk and Slab}
\end{figure}
For the sake of understanding the nature of surface states in detail, let us consider two different geometries: (i) slab where it is fintie along $y$ direction and (ii) NW where it is finite along both $x$- and 
$y$-directions. In both the slab and NW geometry, we analyse the locations of the FAs with their spin-textures. For the NW geometry which is our main concern, we discuss the surface states in 
WSM and WSC phase, with both open boundary condition (OBC) and periodic boundary condition (PBC). 

\subsection{Model Hamiltonian}
We consider an ISB WSM described by the second-quantized Hamiltonian on a cubic lattice with lattice spacing $l$ ($=1$) given by,
\begin{equation}
 \mc{H}=\sum_{\mbf{k}}\psi^\dagger_\mbf{k} \,\mc{H}_{\rm WSM}(\mbf{k})\,\psi_\mbf{k}\ ,
\end{equation}
  where, $\psi_\mbf{k}\!\!=\!(c_{\mbf{k}\!,A,\up\!},c_{\mbf{k}\!,B,\up\!},c_{\mbf{k}\!,A,\dn\!},c_{\mbf{k}\!,B,\dn\!})^T$ and $c_{\mbf{k}\!,\alpha,\sigma\!}~(c^\dagger_{\mbf{k}\!,\alpha,\sigma\!})$
represents the annihilation (creation) operator for an electron in the orbital $\alpha(=A,B)$ and spin $\sigma=(\up,\dn)$. The momentum, $\mbf{k}$ ($=\{k_x,k_y,k_z\}$), run over the first BZ. Here, $\mc{H}_{\rm WSM}(\mbf{k})$ is described by the four band model as~\cite{Kourtis2016,Zhang2018b,Saxena2023},
\begin{eqnarray}
\mc{H}_{\rm WSM}(\mbf{k})\! \!&=&\! \!\lambda_x\sin k_x \tau_1 s_3 \!+ \!\lambda_y \sin k_y \tau_2 s_0 +\! \beta \tau_2 s_2 + \!   \non \\  &&\alpha \sin k_y \tau_1 s_2 \!+\!\,[(m-4t) + 2t\!\!\!\!\sum_{j=x,y,z}\!\!\!\cos k_j] \tau_3 s_0\ ,  \nonumber \\ \label{Eq.ISB WSM Model}
\end{eqnarray}
where, $\lambda_{x,y}$ and $t$ represent the spin-orbit coupling and nearest-neighbour hopping amplitudes, respectively, $m$ is the crystal-field splitting energy, and $\beta,\,\alpha$ are the real parameters of the model. The Pauli matrices $\tau_i$ and $s_i$ for $i\in \{0,1,2,3\}$ act on the orbital and spin degrees of freedom of the electron, respectively. The term $\beta \tau_2s_2$ in Eq.~\eqref{Eq.ISB WSM Model} is responsible for the breaking of IS, whereas the TRS is preserved following the conditions: $\mc{P}^\dagger \mc{H}_{\rm WSM}(\mbf{k}) \mc{P} \ne \mc{H}_{\rm WSM}(-\mbf{k})$ and $
 \mc{T}^\dagger \mc{H}_{\rm WSM}(\mbf{k})\mc{T} = \mc{H}_{\rm WSM}(-\mbf{k})$
where $\mc{P}=\tau_3s_0$ and $\mc{T}=is_2\mc{K}$, with $\mc{K}$ being the complex conjugation operator. 
After diagonalization the Hamiltonian in Eq.~(\ref{Eq.ISB WSM Model}), the eigenvalues for $\mc{H}_{\rm WSM}$ are obtained as,
\begin{eqnarray}
	E^2(\mathbf{ k}) = \left[ \sqrt{\sin^2k_x + \sin^2k_y} \pm \beta\right]^2 + (\alpha \sin k_y)^2 + \non \\ \left(m-4t + 2t\sum_{j=x,y,z} \cos k_j\right)^2. \label{Eq. Bulk_band_struct}
\end{eqnarray}

Broken IS leads to four bulk WNs $Q_s\,(s=1,2,3,4)$ located at $(\pm k_x^0,0,\pm k_z^0)$, with 
$k_x^0 = \sin^{-1}(\beta)$ and $k_z^0= \cos^{-1}(1-\sqrt{1-\beta^2} -\frac{m}{2t})$ as depicted in Fig.~\ref{Fig:Bulk and Slab}(a-b). The presence of TRS implies that the WNs must appear in Kramer pairs (KPs)~\cite{Armitage2018,Rao2016}. In our model, $Q_1,Q_3$ and $Q_2,Q_4$ form these KPs. TRS also ensures that WNs within a KP share the same chirality. Consequently, the overall chirality of the system remains zero. Following that, the chiralities of $Q_2, Q_4$ are opposite to that of $Q_1, Q_3$.

%

For the superconducting part of the Hamltonian, we consider $s$-wave spin-singlet intra-orbital 
pairing which couples the electrons and holes between the WNs with same chirality i.e., between 
$Q_1, Q_3$ and $Q_2, Q_4$~\cite{Zhang2018b,Zhang2018a,Saxena2023}.
With this consideration, the Bogoliubov-de Gennes (BdG) Hamiltonian for the WSC part 
can be written as:
 \begin{equation}
 	\mc{H}_{\rm BdG} =(1/2)\sum_{\mbf{k}}\Psi^\dagger_\mbf{k}
 		\mc{H}_{\rm WSC}(\mbf{k})\Psi_\mbf{k}\ ,
\end{equation}
 	where,
 	\begin{widetext}
 	\begin{eqnarray}
 	\mc{H}_{\rm WSC}(\mbf{k})= \! \lambda_x\sin k_x \tau_1 s_3 \pi_0 + \! \lambda_y \sin k_y \tau_2 s_0\pi_3 + \beta \tau_2 s_2 \pi_3 + \alpha\,\sin k_y \tau_1 s_2 \pi_3 + \Delta\,\tau_0s_2\pi_2 -\mu \tau_0s_0\pi_3  \non \\ + [(m-4t) + 2t\!\!\!\sum_{j=x,y,z}\!\!\cos k_j] \tau_3 s_0\pi_3\ ,  \label{Eq:WSC_Ham}
\end{eqnarray}
 \end{widetext}
with $\Psi_\mbf{k} =(c_{\mbf{k},A,\up},c_{\mbf{k},B,\up},c_{\mbf{k},A,\dn},c_{\mbf{k},B,\dn},c^\dagger_{\mbf{-k},A,\up},c^\dagger_{\mbf{-k},B,\up},\\c^\dagger_{\mbf{-k},A,\dn},c^\dagger_{\mbf{-k},B,\dn})$ as the Nambu spinor. The Pauli matrices, $\pi_i \,(i=0,1,2,3)$, act on the particle-hole degree of freedom. Here, $\Delta$ denotes the $s$-wave pairing potential and $\mu$ is the chemical potential measured with respect to Weyl nodes. For the rest of the article, we choose the following parameter values in our model: $\lambda_x=\lambda_y=1$, $t=1$, $\beta=0.9$, $\alpha=1.0$, $\mu=0$, $m=0$, 
$\Delta=0.5$ and accordingly, we have $k_x^0\simeq1.11$ and $k_z^0 \simeq0.97$. Note that, the qualitative behavior of our results are not sensitive to the change in 
the parameter values as long as WSM phase is preserved.
\begin{figure*}
	\includegraphics[scale=0.4,page=1]{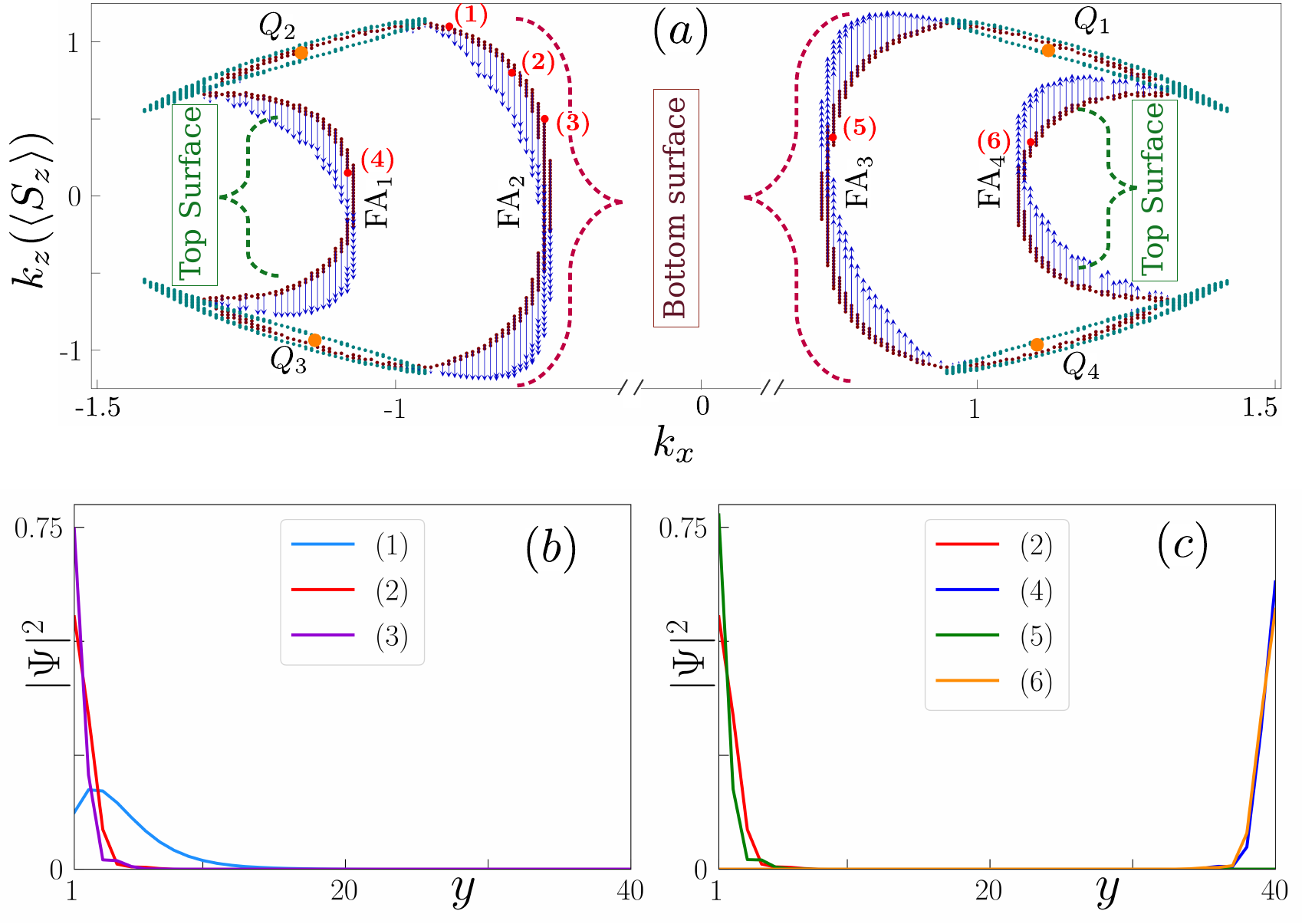}
	\caption{(a) The features of Fermi arcs (FAs) (collection of brown points) for the ISB WSM, labelled as FA$_{\gamma}$ ($\gamma=1,2,3,4$) are depicted along with four WNs, $Q_s$ ($s=1,2,3,4$) (orange color), in $k_x-k_z$ plane considering slab geometry with $L_y=40$ and $E=0.08t$. Bulk Fermi surfaces are shown around each WN in teal color. The spin-polarizations ($\la S_z \ra$ in the $k_x-k_z$ plane) for the states on FAs are shown by blue arrows, where vertically upward (downward) arrows indicate the positive (negative) values of $\la S_z \ra$, respectively, with the length of each arrow being proportional to the magnitude of $\la S_z \ra$. (b\,-\,c) $|\Psi|^2$ is illustrated as a function of position $y$ for the six states (marked by red dots in panel (a)) to highlight their locations.}
	\label{Fig:FAs}
\end{figure*}

\subsection{Slab geometry}

With the discussions on the bulk properties, we now focus on the surface FA states in WSM. For that, we consider a slab geometry schematically shown in Fig.~\ref{Fig:Schematic}(a). Since the WNs are located in the $k_x\!-k_z$ plane, we choose a slab to have a finite size along $y$-direction with thickness $L_y$ such that the projection of all the four WNs can be observed in the sBZ. Along the $x$-and $z$-directions, the WSM slab is infinite, so the momenta along these directions are still good quantum numbers. Consideration of the finite size along the $y$-direction gives rise to two surfaces located at $y=1$ (bottom surface) and $y=L_y$ (top surface) as shown Fig.~\ref{Fig:Schematic}($a$). 

Since the momentum along the $y$-direction is not well defined, to find the band structure in slab geometry we first obtain the real space Hamiltonian by performing an inverse Fourier transformation (FT) only along the $y$-direction using  
\begin{eqnarray}
c_{\mbf{k},\alpha,\sigma} &=& \frac{1}{\sqrt{L_y}}\sum_{y=1}^{L_y} e^{ik_y y}  c_{k_xk_zy,\alpha,\sigma}\ , ~~~~{\rm and} \non \\ 
\psi_\mbf{k}& =&\frac{1}{\sqrt{L_y}}\sum_{y=1}^{L_y} e^{ik_y y} \psi_{k_xk_zy}\ .
\end{eqnarray}
With this transformation, the Hamiltonian takes the form,
 \begin{equation}
  \mc{H}^{\rm slab} = \sum_{k_x\!k_z}\!\!\sum_{\,y,y^\prime\!=\!1}^{L_y} \!\!\psi^\dagger_{k_xk_zy} \mc{H}^{\rm slab}_{\rm WSM}(k_x,k_z,y,y^\prime)\psi_{k_xk_zy^\prime}\ , \non 
\end{equation}
where,
\begin{widetext}
\begin{eqnarray}
 \mc{H}^{\rm slab}_{\rm WSM}(k_x,k_z,y,y^\prime) &=& [\lambda_x \sin k_x \tau_1 s_3 + \beta \tau_2 s_2  + (m-4t)\tau_0 s_0 + \,2t(\cos k_x + \cos k_z)\tau_3 s_0]\delta_{y,y^\prime}  \non \\&& ~~~~~~~~~~~~~~+ [(\lambda_y/2i)\tau_2 s_0 +(\alpha/2i)\tau_1 s_2  + t\tau_3 s_0] \delta_{y,y^\prime +1} + {\rm h.c.}
\end{eqnarray}
\end{widetext}
We numerically diagonalize the above Hamiltonian for each set of $(k_x, k_z)$ values with $L_y=40$ slices considering both OBC and PBC along $y$-direction. Note that, surface states can be observed only in OBC, while PBC is employed to identify only the bulk states. We show the band structure as a function of $k_x$ by setting $k_z = 0.9$ in Fig.~\ref{Fig:Bulk and Slab}(c). The choice of $k_z$ can be made anywhere in between [$-k_z^0,k_z^0$] since the surface states only appear in between the WNs with opposite chiralities. Similarly, we depict the band structure as a function of $k_z$ with $k_x =1$ 
in Fig.~\ref{Fig:Bulk and Slab}(d) employing both OBC and PBC along $y$-direction. In both the figures, we observe the gapless dispersive FA surface states between the WNs when OBC is implemented.

Now, we investigate the shape of the FAs obtained in the WSM slab geometry. FAs are the open constant energy contours in the $k_x-k_z$ plane residing at the boundary of the system. We fix the energy at $E=0.08t$ ($t$ sets the energy scale in the system) and draw the Fermi surface in Fig.~\ref{Fig:FAs}($a$). With the OBC, we observe the existence of four FAs denoted by FA$_\gamma$ 
with $\gamma=1,2,3,4$. As expected, two FAs connect WNs with opposite chirality i.e., FA$_1$ and FA$_2$ (FA$_3$ and FA$_4$) connect the WNs, $Q_2$ and $Q_3$ ($Q_1$ and $Q_4$). On top of FAs, 
we also show the Fermi surface in PBC, which only contains the bulk states around each WN as shown in Fig.~\ref{Fig:FAs}($a$) using teal color. Interestingly, the shape of FAs obtained in this model closely resembles the FAs observed in real materials~\cite{Lv2015c,Xu2016,Sun2015,Xu2015a,Xu2015c,Lv2015b}.

To get insight about the locations of the states on the FAs in the real space, we depict $|\Psi|^2$ 
($\equiv |\psi_{k_xk_z}(y)|^2$) as a function of $y$ in Fig.~\ref{Fig:FAs}(b-c) for the states on the 
FAs marked in red dots [see Fig.~\ref{Fig:FAs}($a$)]. We observe that the FA states are either localized on the top surface ($y=L_y$) or at the bottom surface ($y=1$). Specifically, the states in the FA$_2$ are localized at the bottom surface (see Fig.~\ref{Fig:FAs}(b)). We also note that, on FA$_2$ the states away from the WNs i.e., close to the centre of the arc, are distinctly localized at the bottom surface (see curve~2 and 3 in Fig.~\ref{Fig:FAs}(b)). On the other hand, the states close to the WNs have significant overlap with the bulk states (see curve~1 in Fig.~\ref{Fig:FAs}(b)). This happens since the FAs leak into the bulk states near the WNs as mentioned in the previous section. Similarly, we also choose points from FA$_1$, FA$_3$, FA$_4$ (red colored dots marked by 4, 5, 6 in Fig.\,\ref{Fig:FAs}(a)) and present the behavior of $|\Psi|^2$ in Fig.\,\ref{Fig:FAs}(c). We find that FA$_1$ and FA$_4$ states are localized on the top surface, while those for FA$_2$ and FA$_3$ are localized at the bottom surface. Note that, we show the curve~2 in panel (b) too for the sake of comparison and clarity.


Here, we explore the spin textures of the FA states. We compute the expectation value of the spin operator along $z$-direction for each sites along the $y$-direction. To perform that, we expand the states in terms of the basis $\ket{y,\alpha,\sigma}$ with $y=1,2,...L_y; \sigma=\up,\dn; \alpha=A,B$ 
for the slab geometry as
\begin{equation}
\ket{\psi_{k_xk_zy}^{\rm FA}} = \sum_{y=1}^{L_y}\!\sum_{\sigma=\up\!,\dn}\!\sum_{\alpha\!=\!A\!,B\!} d(y,\alpha,\sigma)\ket{y,\alpha,\sigma}\ .
\end{equation}
The spin-polarization along $z$-direction is defined as 
\begin{equation}
S_z=\frac{1}{2} (\ket{\up}\bra{\up}-\ket{\dn}\bra{\dn})\ ,
\end{equation}
and the corresponding expectation value of $S_z$ is given by
\begin{eqnarray}
\bra{\psi_{k_xk_zy}^{\rm FA}}\!S_z\!\ket{\psi_{k_xk_zy}^{\rm FA}}\!=\! \sum_{y=1}^{L_y}\sum_{\alpha=A,B}\! \frac{1}{2}(|d(y,\alpha,\up)|^2-\!|d(y,\alpha,\dn)|^2)\ . \non \\
\end{eqnarray}

We show the spin-textures i.e., the expectation values $\la S_z \ra$ taking into account the FA states
in Fig.~\ref{Fig:FAs}(a) for the sake of understanding. The up (down) arrows, $\up$($\dn$), are used to express the positive (negative) values of $\la S_z \ra$. The length of each arrow is proportional to the value of $\la S_z \ra$ with a maximum value of $1/2$. We observe that the states on FAs exhibit both up and down spin-polarization. Specifically, the states on FA$_1$ and FA$_2$ (FA$_3$ and FA$_4$) host down (up) spin-polarized states. Now, focussing on FA$_1$ and FA$_4$, we infer that the electrons localized on the top surface have spin polarizations along both positive and negative $z$-axis. Notably, this information is very crucial for the formation of superconducting pair between the surface state electrons indicating a strong possibility of Andreev reflection in WSM-WSC hybrid junction mediated 
by the surface states. Similarly, the FA$_2$ and FA$_3$ states are localized at the bottom surface and contain both spin-polarizations. Note that, $|\la S_z \ra| \ne 1/2$ for all the states on FAs. States in the centre of FAs have $|\la S_z \ra| = 1/2$ while states close to the WNs are partially spin polarized. The presence of TRS in the system is also reflected in the spin textures of FAs since any FA state with $(k_x,k_z)$ and its time-reversed partner $(-k_x,-k_z)$ have spin polarizations opposite to each other.
\begin{figure}
	\includegraphics[scale=0.35,page=1]{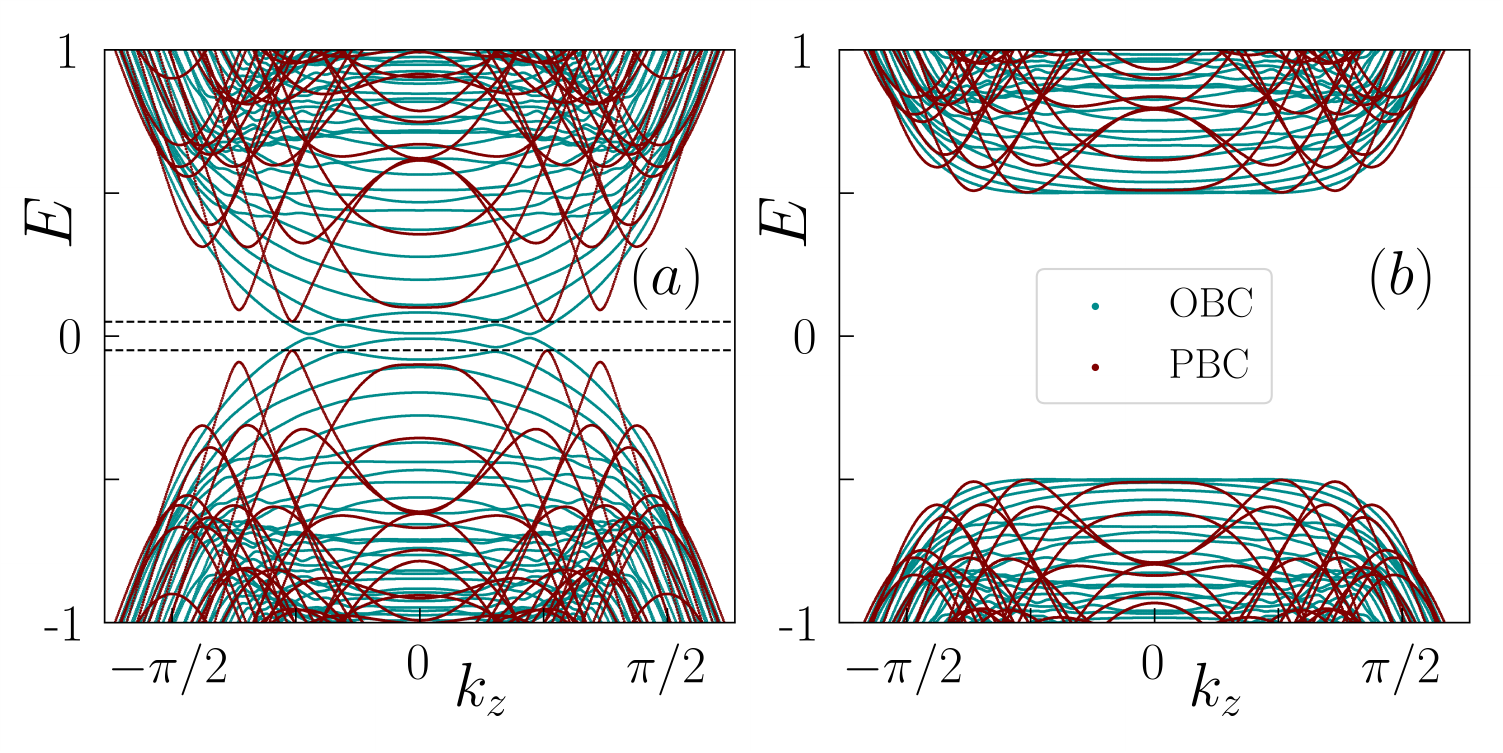}
	\caption{Band structure is shown as a function of $k_z$ for ($a$) WSM and (b) WSC NW choosing $L_x=L_y=20$ and employing both OBC (teal color) and PBC (dark-red color) along $x$ and $y$ directions. The finite size gap, $E_g~(=0.05t)$, in WSM NW bulk band structure is highlighted by the dashed lines.}
	\label{Fig:NW_band}
\end{figure}
\vspace {0.5cm}
\subsection{Nanowire geometry}

With the understanding of the surface FA states in WSM slab geometry, we now turn our focus to the NW geometry, which is the prime concern of our present work. The size of the NW along $x$ and $y$-directions are $L_x$ and $L_y$, respectively, where, $L_x=L_y$. To find the band structure of the NW, we consider the NW to be infinite along $z$-direction so that the momentum along $z$-direction becomes well defined (PBC), while the momenta along $x$ and $y$-direction remain ill-defined (OBC) due to the finite size.
\subsubsection{{\rm{WSM NW}}}
For the WSM NW, we obtain the Hamiltonian by performing the inverse FT along $x$ and $y$-directions using 
\begin{eqnarray}
c_{\mbf{k},\alpha,\sigma} &=& \frac{1}{\sqrt{L_x L_y}}\sum_{x=1}^{L_x}\sum_{y=1}^{L_y} e^{i(k_x x + k_y y) } c_{x\, y\, k_z,\alpha,\sigma}\ , {\rm ~~~and} \non \\
\psi_\mbf{k} &=& \frac{1}{\sqrt{L_x L_y}}\sum_{x=1}^{L_x}\sum_{y=1}^{L_y} e^{i(k_x x + k_y y) } \psi_{x\, y\, k_z,\alpha,\sigma}.
\end{eqnarray}
With this transformation, the Hamiltonian takes the form, 
 \begin{equation}
\mc{H}^{\rm NW} = \sum_{k_z}\sum_{x,x^\prime=1}^{L_x} \sum_{y,y^\prime=1}^{L_y} \psi^\dagger_{x\,y\,k_z} \mc{H}^{\rm NW}_{\rm WSM}(x,x^\prime, y,y^\prime,k_z) \psi_{x^\prime,y^\prime,k_z} \non 
\end{equation}
where,
\begin{widetext}
\begin{eqnarray}
\mc{H}^{\rm NW}_{\rm WSM}(x,x^\prime,y,y^\prime,k_z) &=& \left(\frac{\lambda_x}{2i}\tau_1 s_3 +t \tau_3 s_2\right)\delta_{x,x^\prime +1} \delta_{y,y^\prime} + [\beta \tau_2 s_2 +  (m-4t)\tau_0 s_0 +  \cos k_z\tau_3 s_0] \delta_{x,x^\prime}\delta_{y,y^\prime} \non \\ && ~~~~~~~~~~~~~~~~~~~~~~~~~~~~~~~~~+\left[\frac{\lambda_y}{2i}\tau_2 s_0 +\frac{\alpha}{2i} \tau_1 s_2  + t\tau_3 s_0\right] \delta_{x,x^\prime}\delta_{y,y^\prime +1} + {\rm h.c.} \ , \label{Eq. WSMNW Ham}
\end{eqnarray}
\end{widetext}
For the WSM NW, we consider $L_x=L_y=W=20$ and numerically diagonalize the above Hamiltonian for each value of $k_z$. We present the eigen spectrum as a function of $k_z$ in Fig.~\ref{Fig:NW_band}(a) employing both PBC and OBC. As mentioned earlier, in OBC,` we obtain the information about both bulk and surface states of the system while in PBC information about only the bulk states can be achieved.

For TRS-broken WSM NW, it has already been shown in Ref.~[\onlinecite{Kaladzhyan2019}] that the bulk states are gapped out due to the finite size effect, and within the bulk gap, $E_g^{\rm bulk}$, only surface states exist. A finite size gap, $E_g^{\rm surface}$, is also developed on the surface state spectrum but the bulk confinement gap is larger than that of developed in the surface. Specifically, $E_g^{\rm bulk} \sim 1/W$ whereas $E_g^{\rm surface} \sim 1/4W$. This feature is also observed in the present model where the bulk gap $E_g=0.05t$ and $E_g^{\rm bulk}\sim 1/W$ and within the energy regime, $[-E_g,E_g]$, only surface states are present [shown by teal color in Fig.~\ref{Fig:NW_band}(a)]. This behaviour is not present in the slab geometry where both the bulk and surface states are gapless, and thus distinguishing the surface states from the bulk do not seem possible in the WSM slab. Therefore NW geometry is the possible platform where the surface states can be clearly distinguished from the bulk states and can possibly be probed in such a way that the contribution of the bulk states in the measurement can be excluded. 

\begin{figure}
	\includegraphics[scale=0.3]{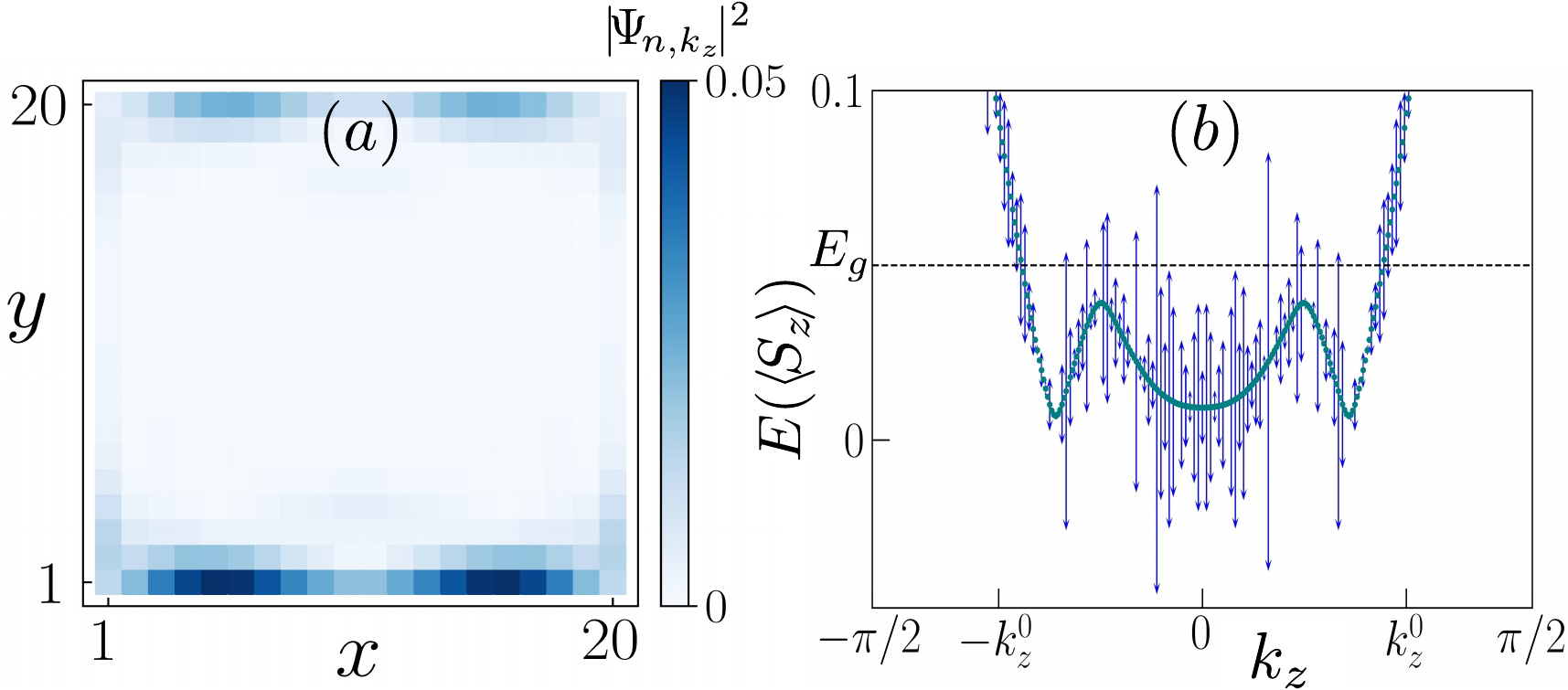}
	\caption{$(a)$ Probability of wave-function $|\Psi_{n,k_z}|^2$ is shown as a function of the position $x$ and $y$ with $k_z=0$ for the state corresponding to the lowest positive energy band 
	within the gap $E_g$ as illustrated in Fig.~\ref{Fig:NW_band}$(a)$. $(b)$ Average spin polarization $\la S_z \ra$ for those states are depicted by the blue arrows on top of the lowest positive energy 
	band (teal color). The direction and length of all arrows carry the same meaning as mentioned in Fig.\,\ref{Fig:FAs} .}
	\label{Fig:NW_SpinTex}
\end{figure}
Furthermore, similar to the slab geometry, we here discuss the properties of FAs in the NW geometry. In NW geometry, FAs are present within the bulk confinement gap, \ie $-E_g\!\le \!E\!\le\! E_g$. 
From the band structure, presented in Fig.~\ref{Fig:NW_band}$(a)$ (shown by teal color), it is clear that only the lowest energy band with $E\gtr0$ is present inside the confinement gap, $E_g$, between 
the WNs ($-k_{z}^0\le k_z\le k_{z}^0$). Therefore, we focus on this band to obtain the properties of FAs. First, we compute $|\Psi_{n,k_z}(x,y)|^2$ as a function of position, $(x,y)$, with $k_z=0$ where 
$\Psi_{n,k_z}$ is the eigenstate of the Hamiltonian [Eq.~\eqref{Eq. WSMNW Ham}] corresponding to the lowest positive energy band having momentum, $k_z$. We depict the probability $|\Psi_{n,k_z}(x,y)|^2$ in Fig.~\ref{Fig:NW_SpinTex}$(a)$ and establish that the states within the bulk gap, $E_g$, are indeed localized on the surface of the NW with significant population in $y=1$ and $y=L_y(=20)$ surface. Then, we compute the expectation value of polarization along $z$-axis in the earlier mentioned energy band and show it in Fig.~\ref{Fig:NW_SpinTex}$(b)$. We observe that the states within this band, 
carry polarization along both positive and negative $z$-direction. Presence of both up and down spin polarized states on the NW surface, is very crucial for generating AR process mediated via FAs in a WSM/WSC NW junction [see Fig.\ref{Fig:Schematic}(b))] which we discuss elaborately in Sec.~\ref{Sec:III} B.
\subsubsection{{\rm{WSC NW}}}
After explaining the nature of FAs in WSM NW, let us discuss the effect of superconductivity in the NW geometry. We consider an $s$-wave spin-singlet intra-orbital pairing with amplitude $\Delta$ in both the bulk and surface of the NW. Since the surface states contain electrons with both spin polarizations (as shown in the slab geometry calculations), $s$-wave spin-singlet pairing between the electrons is expected to be prominent over the spin-triplet pairing. Similar to the WSM NW, we perform the inverse FT to obtain the WSC NW Hamiltonian as,
\begin{widetext}
	\begin{eqnarray}
	\mc{H}^{\rm NW}_{\rm WSC}(x\!,x^\prime\!,y,y^\prime\!,k_z\!) &=& \left(\frac{\lambda_x}{2i}\tau_1 s_3 \pi_0 +t \tau_3 s_2 \pi_3\right)\delta_{x,x^\prime +1} \delta_{y,y^\prime} + [\beta \tau_2 s_2 \pi_3 + (m-4t)\tau_0 s_0 \pi_3 +  \cos k_z\tau_3 s_0\pi_3] \delta_{x,x^\prime}\delta_{y,y^\prime}  \non \\ &&~~~~~~~~~~~~~~~+ \left[\frac{\lambda_y}{2i}\tau_2 s_0 +  \frac{\alpha}{2i} \tau_1 s_2 \pi_3  +t\tau_3 s_0\right] \delta_{x,x^\prime}\delta_{y,y^\prime +1} +   \Delta \delta_{x,x^\prime}\delta_{y,y^\prime} \tau_0s_2\pi_2  + {\rm h.c.}\ , \label{Eq.WSC NW Ham}
	\end{eqnarray}
\end{widetext}
We then numerically diagonalize the Hamiltonian considering the same system size as mentioned for the WSM NW and plot the band structure as a function of $k_z$ in Fig.~\ref{Fig:NW_band}(b) employing both OBC and PBC. We observe that due to the superconducting correlation, both bulk and surface states acquire a gap of magnitude $\Delta$. 

\section{Conductance}\label{Sec:III}

In this section, we present our numerical results for the conductance in WSM and WSM/WSC NW setup.
\begin{figure*}
	\includegraphics[scale=0.52,page=1]{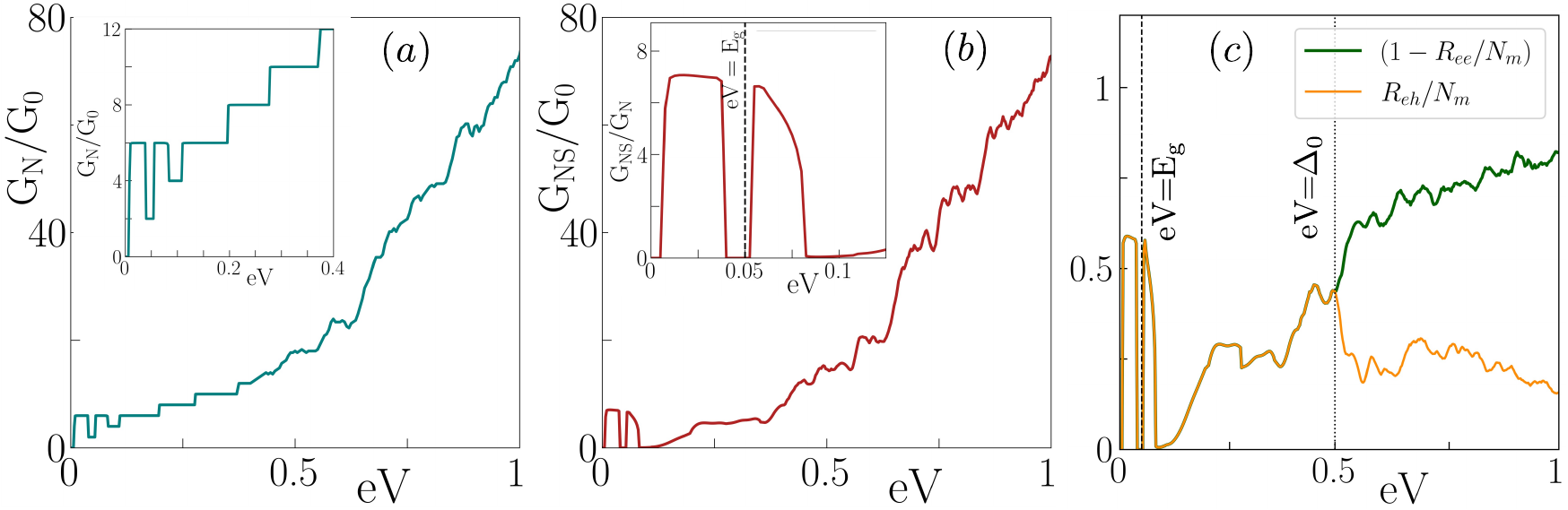}
	\caption{Two-terminal conductance in units of quantum conductance, $G_0 = (e^2/h)$ is illustrated as a function of voltage bias $eV$ in ($a$) WSM NW and ($b-c$) WSM/WSC NW hybrid junction 
	considering $L_x=L_y=20$ and $L_z=80$ lattice sites. ($c$) $(1-R_{ee}/N_m)$ and $R_{eh}/N_m$ are depicted as a function voltage bias $eV$, explicitly exhibiting the contribution to conductance 
	via AR process within the bulk confinement gap ($eV<E_g$) and superconducting gap $(eV<\Delta_0)$.}
	\label{Fig:NWConductance}
\end{figure*}
\subsection{WSM NW}
Let us begin by analyzing the transport signatures of FAs in an ISB WSM NW based two-terminal setup. For this purpose, we first exclude the WSC NW part in Fig.~\ref{Fig:Schematic}(b) by extending the WSM NW and attach two semi-infinite leads at $z=-L/2$ and $z=L/2$. We model both the leads using the same Hamiltonian which is used to describe the WSM NW. The chemical potential at the left (right) lead is fixed at $\mu_L$~($\mu_R$). Under the application of voltage bias, $eV (=\mu_L-\mu_R)$, we compute two-terminal charge transport employing Landauer formula~\cite{datta_1995}. To obtain the current traversing through the NW, we first construct the scattering matrix, which relates the incoming propagating modes to the outgoing modes in the leads with the central WSM NW being considered 
as the scatterer. Incoming, outgoing states and the scattering matrix are defined as:

\begin{eqnarray}
\Psi^{\rm in}=& [\psi_1^{\rm L},\psi_2^{\rm L},...,\psi_{4N_L}^{\rm L},\psi_1^{\rm R},\psi_2^{\rm R},...,\psi_{4N_R}^{\rm R}]^T\ , \\ 
\Phi^{\rm out}=&[\phi_1^{\rm L},\phi_2^{\rm L},...,\phi_{4N_L}^{\rm L},\phi_1^{\rm R},\phi_2^{\rm R},...,\phi_{4N_R}^{\rm R}]^T \ ,
\end{eqnarray}
\begin{center}
$\Phi^{\rm out} = \hat{S}\Psi^{\rm in}$\ ,
\end{center}
where, $\psi^{\rm L(R)}_i$ is the incoming state from the left (right) lead in the $i^{\rm{th}}$ mode. Here, $4N_L(4N_R)$ is the number of occupied modes/channels in the left (right) lead for a given voltage bias $eV$ (including both spin and orbital degrees of freedom). Similarly, $\phi^{\rm L(R)}_i$ is the outgoing state into the left (right) lead in the $i^{\rm{th}}$ mode after the scattering event takes place. Here, `$T$' denotes the tranpose operation. 
The unitary scattering matrix $\hat{S}$ of dimension $(4N_L+4N_R)\times(4N_L+4N_R)$ reads
 \begin{equation}
 \hat{S}=
 \begin{bmatrix}
 \hat{r} & \hat{t^\prime} \\
 \hat{t} & \hat{r^\prime} \ 
 \end{bmatrix}\ ,
 \end{equation}
where, $\hat{r} \,(\hat{r^\prime})$ is a square matrix of dimension $4N_L\times 4N_L(4N_R\times 4N_R)$ and $\hat{t}\, (\hat{t^\prime})$ is a matrix of dimension $4N_R\times 4N_L(4N_L\times 4N_R)$. Physically, $\hat{r}\,(\hat{r^\prime})$ represents the reflection matrix with elements $r_{i,j}\,(r^\prime_{i,j})$ being the amplitude of reflection from the $j^{\rm{th}}$ mode to the $i^{\rm{th}}$-mode in the left (right) lead. Similarly, $\hat{t}\,(\hat{t^\prime})$ represents the transmission matrix with elements $t_{i,j}\,(t^\prime_{i,j})$ denoting the amplitude of the transmission from the $j^{\rm{th}}$ mode in left (right) lead 
to the $i^{\rm{th}}$ mode in the right (left) lead following the unitarity condition: $r^\dagger r + t^\dagger t = \mc{I}$. Within this formalism, the two-terminal conductance at zero temperature can be obtained using the Landauer formula given by~\cite{datta_1995},
	\begin{equation}
		G_{\rm N} \,(eV)=G_0 \, \text{Tr}[t^\dagger t]|_{E=eV}\ , 
		\label{Eq.Landauer Formula}
	\end{equation}
where, $G_0=e^2/h$ is the unit of quantum conductance. The scattering amplitudes can be calculated numerically using python package KWANT~\cite{Groth2014}.

We depict the two-terminal conductance in WSM NW setup in units of quantum conductance as a function of voltage bias, $eV$, in Fig.~\ref{Fig:NWConductance}(a) assuming the length along $z$-direction is $L_z=80$ lattice sites. We observe that conductance initially increases in steps of $G_0$ and a plateau-like behaviour appears after that. For a more clear understanding we refer to the inset of Fig.~\ref{Fig:NWConductance}(a). However, when we increase the bias voltage, the step-like behaviour is lost and a continuous enhancement in conductance is observed. Similar behaviour of the conductance has also been observed for TRS-broken WSM NW~\cite{Kaladzhyan2019}. Interestingly, in case of ISB WSM NW, the quantization of the conductance occurs in steps of $2G_0$, whereas in case of TRS-broken WSM NW, the steps appear in units of $G_0$ as shown in Ref.~[\onlinecite{Kaladzhyan2019}]. This happens due to the presence of four FAs (two on each surface) in such system as compared to TRS-broken WSM where two FAs are present. The step-like behaviour appears due to the FA surface states within the finite size gap in the system, while the following continuous enhancement happens since the bulk states start contributing. Due to semimetallic nature of the bulk spectrum, the conductance for the higher bias voltages varies as $(eV)^2$. 

In clean systems, the conductance does not depend on the  length of the WSM NW along the $z$-direction because of phase coherent transport along the NW axis considering the same Hamiltonian 
for the leads. Increasing the system size along the transverse directions do not affect the qualitative picture of the conductance plot, but the quantitative behaviour of the conductance changes since 
both $E_g^{\rm bulk} (\sim 1/W)$ and $E_g^{\rm surface} (\sim 1/4W)$ decrease. Additionally, increasing the value of $L_x$ and $L_y$ usually enhances the number of transverse modes in the lead spectrum. Hence, the occupancy of propagating modes increases within the leads for a given voltage bias, which in turn can enhance the conductance according to Eq.~\eqref{Eq.Landauer Formula}.

\subsection{WSM/WSC NW}
Here, we discuss another main finding of our analysis which deals with the transport signatures of the FAs in WSM/WSC NW hybrid junction. For this purpose, we consider the geometry shown in 
Fig.~\ref{Fig:Schematic}(b) under the application of voltage bias $eV$. We model this hybrid setup using the Hamiltonian in Eq.~\eqref{Eq.WSC NW Ham}. 
The uniform superconducting pairing potential is chosen as,
\beq
\Delta(x,y,z)=
\left\{
\begin{aligned}
 &\Delta_0 ~~ \forall z\gtr 0 \,,\, x,y \in (0,W)  \\
			  &0 ~~~~\,\forall z< 0\,,\, x,y \in (0,W) \ .
\end{aligned}
\right.
\eeq
The leads are also modelled by the same Hamiltonian as mentioned in Eq.~\eqref{Eq.WSC NW Ham}. The left lead is chosen to be nonsuperconducting ($\Delta_{0}=0$), while the right lead possesses 
a superconding pairing gap as mentioned above. This hybrid setup mimics a normal-superconductor (NS) junction. 

The additional mechanism that comes into play while considering charge transport in such superconducting hybrid junction, is the Andreev reflection (AR) where an incoming right moving electron 
from the left lead with spin $\sigma$ combines with another electron with oppsosite spin $\bar{\sigma}$ to form a spin-singlet Cooper pair (CP) leaving behind a hole that reflects back from the interface. The CPs, formed at the interface, propagates through the WSC and give rises to supercurrent~\cite{Blonder1982}. In our work, the primary motivation to capture the signatures of FAs via the AR lies into these spin textures of the FAs [see Fig.~\ref{Fig:FAs}(a)-(c)]. Note that, within a particular surface, the spin polarization of electrons along the $z$-direction has components along both positive and negative $z$-axis, indicating the strong possibility of AR mediated via the FAs.

To find the conductance in this hybrid junction, we employ the scattering matrix formalism which now takes more complex form compared to the bare WSM NW due to the presence of the AR process. 
It can now be written as
\begin{equation}
\begin{bmatrix}
\Phi^{\rm L}_e \\  \Phi^{\rm L}_h \\ \Phi^{\rm R}_e \\ \Phi^{\rm R}_h
\end{bmatrix} = \mathcal{S}_{\rm NS} 
\begin{bmatrix}
\Psi^{\rm L}_e \\  \Psi^{\rm L}_h \\ \Psi^{\rm R}_e \\ \Psi^{\rm R}_h
\end{bmatrix}\ ,
\end{equation}
where, $\Psi^{\rm L(R)}_{e(h)}$ is a column matrix of dimension $4 N_{\rm L(R)}$ and it represents the incoming electron (hole) from the left (right) lead. Similarly, $\Phi^{\rm L(R)}_{e(h)}$ designates 
the outgoing electron (hole) in the left (right) lead. The scattering matrix $\mc{S}_{\rm NS}$ can be written as
\begin{equation}
\mc{S}_{\rm NS}=
\begin{bmatrix}
r_{ee} & r_{eh} & t^\prime_{ee} & t^\prime_{eh} \\
r_{he} & r_{hh} & t^\prime_{he} & t^\prime_{hh} \\
t_{ee} & t_{eh} & r^\prime_{ee} & r^\prime_{eh}  \\
t_{he} & t_{hh} & r^\prime_{he} & r^\prime_{hh} 
\end{bmatrix} \ ,
\end{equation}
where, $r_{ee},r_{eh},r_{he} $ and $r_{hh}$ denote the complex matrices with dimension $4\,N_L \times 4N_L$. The matrix element $(r_{\alpha\beta})_{ij}$ represents the amplitude of reflection from the particle type-$\beta$ in the $j^{\rm{th}}$ channel of left lead to the particle type-$\alpha$ in the $i^{\rm{th}}$ channel of left lead with $\alpha,\beta=(e,h)$. For our purpose, it is now sufficient to focus on the reflection matrices in the left lead i.e.,  
\begin{equation}
\mc{R}_{\rm NS}=
\begin{bmatrix}
r_{ee} & r_{eh}  \\
r_{he} & r_{hh}
\end{bmatrix} \ .
\label{Eq.R_NS}
\end{equation} 
The reason behind writing Eq.~(\ref{Eq.R_NS}) is the absence of the quasiparticle states within the superconducting gap which prevents the transmission of electron like (or hole like) particles from the left normal lead to the right superconducting lead. Similarly, there are no single particle states present inside the right lead to propagate through the WSC and reach the left lead. In this circumstance, within the subgap regime, the only possible scattering processes are: reflection of electrons as an electron (normal reflection), denoted by $r_{ee}$ matrix in Eq.~\eqref{Eq.R_NS}, and AR. Note that, it follows 
the unitarity relation of $\mc{R}_{\rm NS}$ as given by
\begin{equation}
 r_{ee}^\dagger r_{ee} + r_{he}^\dagger r_{he}=\mc{I} \label{Eq.Unitarity relation}\ .
\end{equation} 

With this understanding, we now employ the BTK formalism to obtain the conductance of this hybrid junction as~\cite{Blonder1982,Dumitrescu2015,Rosdahl2018,Das2016}
\begin{equation}
G_{\rm NS}(eV)= G_0 \, [N_m(E) - R_{ee}(E) + R_{eh}(E)]|_{E=eV}\ ,  \label{Eq.BTK}
\end{equation}
where, $R_{ee}(E)=\text{Tr}(r_{ee}^\dagger r_{ee})$ and $R_{eh}(E)=\text{Tr}(r_{he}^\dagger r_{he})$. $N_m(eV)$ is the number of occupied modes/channels in the left lead for a given voltage bias $eV$. Using the python package KWANT~\cite{Groth2014}, we obtain the reflection matrices, $r_{ee}\,,r_{eh}$, and also $N_m$ to compute the conductance. 
\begin{figure}
\hspace{-1.0cm}
	\includegraphics[scale=0.33,page=1]{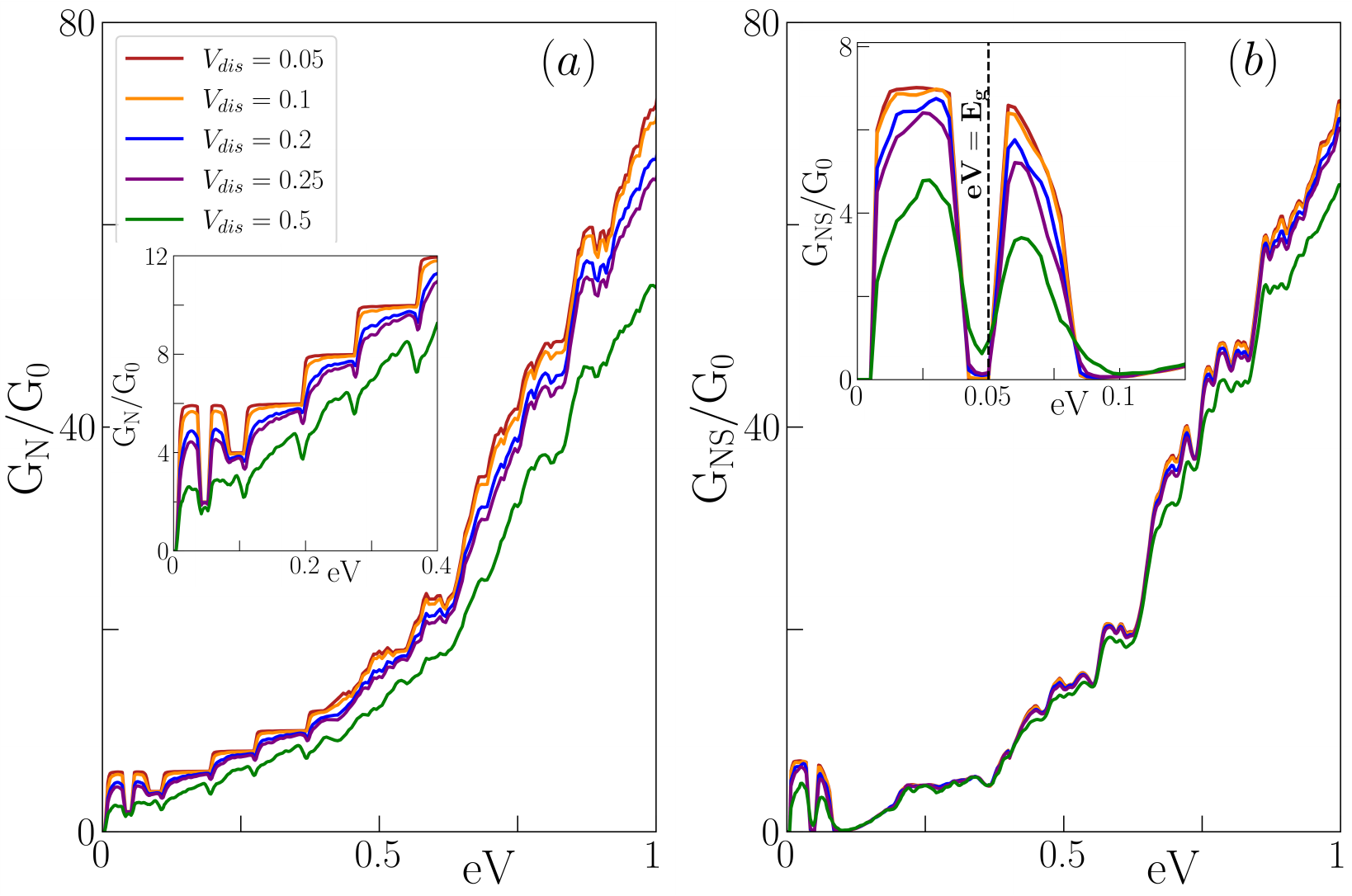}
	\caption{Two-terminal conductance is demonstrated for a disordered (a) WSM NW and (b) WSM/WSC NW junction as a function of voltage bias, $eV$, choosing various disorder strengths, $V_{\rm dis}$. Dimensions of the NWs are considered to be the same as mentioned in Fig.~\ref{Fig:NWConductance}. We choose the delta-correlated disorder to be uniformly distributed between $[-V_{\rm dis}/2,V_{\rm dis}/2]$. We consider 30 disorder configurations for our analysis.}
\label{Fig:Conductance_disorder}
\end{figure}

We depict the conductance, $G_{\rm NS}$, as a function of voltage bias, $eV$, in Fig.\,\ref{Fig:NWConductance}(b). We also show $(N_m-R_{ee})$ and $R_{eh}$ for comparison, normalized by the number of available modes in left lead ($N_m$), as a function of $eV$ in Fig.~\ref{Fig:NWConductance}(c). For the  bias voltage less than the bulk confinement gap ($eV<E_g$) where only the FA surface states exist [see Fig.~\ref{Fig:NW_band}(a)], we find $G_{\rm NS}$ exhibits nonzero value and becomes almost equal to $7G_0$ [see Fig.~\ref{Fig:NWConductance}(b)]. 
This is concomitant with nonzero value of $R_{eh}$ 
[see Fig.~\ref{Fig:NWConductance}(c)] which establishes the possibility of AR mediated by FAs 
which is one of the main claims of the present work: capturing the signatures of FAs in this superconducting hybrid junction. In the subgapped regime ($eV<\Delta_0$), $(N_m-R_{ee})/N_m$ and $R_{eh}/N_m$ identically follow each other which can be explained using the unitarity relation of $\mc{R}_{\rm NS}$ [see Eq.~\eqref{Eq.Unitarity relation}] as, $R_{eh}=\text{Tr}(r_{he}^\dagger r_{he}) =  \text{Tr}(\mc{I} - r_{ee}^\dagger r_{ee}) = N_m - R_{ee}$. For $eV>E_g$, we observe $G_{\rm NS}$ suddenly drops to zero and then gradually increases with the voltage bias as shown in Fig.~\ref{Fig:NWConductance}(b). For better clarity, we also refer to the inset of Fig.~\ref{Fig:NWConductance}(b). When $eV>\Delta_0$, $R_{eh}\neq(N_m-R_{ee})$ and $R_{eh}$ decays with the increase in voltage bias. This is due to the presence of finite quasiparticle density of states for $eV>\Delta_0$, which allows $t_{ee}$ and $t_{he}$ matrices to be nonzero, and as a result, 
Eq.~\eqref{Eq.Unitarity relation} does not hold. 

Interestingly, in the subgapped regime, $G_{\rm NS}/G_0$ is not quantized in the regime where $G_{\rm N}/G_0$ exhibits quantized values as shown in the inset of Fig.~\ref{Fig:NWConductance}(a). 
This is unusual since in the absence of any interfacial insulating barrier (transparnent limit), the subgap condutance is expected to be twice of the conductance in the absence of the superconductors i.e., $G_{\rm NS}=2G_{\rm N}$ (see Ref.~[\onlinecite{Blonder1982}] for details). This pecularity can originate from two correlated reasons as follows. First: even though FAs host both up and down spin textures, 
the expectation value of spin polarization, $|\langle S_z \rangle| \neq 1/2$, for all states on the FAs. This makes the AR deviated from the unit probability within the subgap regime even in the transparent limit 
which indicates that all the electrons may not reflect as holes from the interface. Thus, perfect AR does not take place restricting the quantization of $G_{\rm NS}$. Second: In Fig.~\ref{Fig:NWConductance}(c), we note that a finite value of $R_{ee}/N_m$ (since $N_m$ is quantized in the region of concern) even in the absence of any insulating barrier at the interface. Such normal reflection probability can originate from the inter-channel scatterings in the WSM NW which can also prohibit the perfect quantization.

\section{Stability against disorder}\label{Sec:IV}
So far, all the results are presented for clean systems. We now extend our analysis to include the effect of disorder and investigate the robustness of our results in both WSM NW and WSM/WSC NW junction. Usually, the bulk properties of WSM are robust against disorder unless the disorder strength is strong enough to allow inter-node scatterings and create a gap to destroy the topological phase~\cite{Shapourian2016,Klier2019,Chen2015,Altland2016}. Specifically, to check the stability of FAs against the disorder that breaks translational symmetry of the system, we consider random quenched disorder which are delta-correlated, in terms of a onsite energy potentials in the Hamiltonian as,
\begin{equation}
\mc{H}_{\rm dis}(\mbf{r})= V(r)\Gamma\ ,
\end{equation}
where, $V(\rm r)$ is random number uniformly distributed in the range $[-V_{\rm dis}/2,V_{\rm dis}/2]$ and $V_{\rm dis}$ is referred as the disorder strength. We choose $\Gamma=\tau_0\,s_0$ for WSM NW and $\Gamma=\tau_0\,s_0\pi_3$ for WSM/WSC NW junction. For the hybrid junction, the disorder is considered only in $-L/2<z<0$ region~\cite{Beenakker1992,Takagaki1996,Pientka2012}.

To compute the conductance, $G_{\rm N}$ and $G_{\rm NS}$ (Eq.~\eqref{Eq.Landauer Formula} and Eq.~\eqref{Eq.BTK}), in the presence of disorder, we again employ scattering matrix formalism and extract the reflection ($r_{ee},r_{he}$) and transmission ($t$) matrices using KWANT~\cite{Groth2014}. We show the disorder averaged conductance, $G_N$ and $G_{\rm NS}$, as a function of voltage bias for various disorder strengths in Fig.~\ref{Fig:Conductance_disorder}($a$) and ($b$) respectively. The results are obtained after averaging over 30 disorder configurations.  

The average energy level spacing of the FAs in the WSM NW geometry is estimated approximately as 
$\Delta E_l =0.08t$ where the average level broadening induced by disorder is, $\Delta E_{\rm dis} =\frac{\pi}{3} V_{\rm dis}^2$~\cite{Kaladzhyan2019}. From here, we can estimate the critical disorder strength above which quantized conductance plateau does not survive as, $V_{\rm dis}^{\rm c}=\sqrt{3\Delta E_l/\pi} = 0.27$. From Fig.~\ref{Fig:Conductance_disorder}(a) and Fig.~\ref{Fig:Conductance_disorder}(b), we can verify that quantized conductance plateau, arising due to the FA surface states, survive upto the disorder strength, $V_{\rm dis}=0.25$, (see the inset of Fig.~\ref{Fig:Conductance_disorder}(a) for better clarity) above which the conductance quantization is 
diminished. 
Similarly, in the case of the WSM/WSC NW junction, we observe our results to sustain upto sufficiently large disorder strength. Note that, unlike the clean system, where the conductance is independent of NW length, in the presence of disorder, conductance depends on the NW length since phase coherency is lost. In particular, keeping the disorder strength moderate, the conductance decreases as the NW length is increased.
 
\section{Summary and Conclusions}\label{Sec:V}
To summarize, in this article, we have explored an ISB WSM with four bulk Weyl nodes in $k_x-k_z$ plane. We have analyzed the properties of FAs in a slab geometry considering $y$-direction to be finite and found both up and down spin-polarized FA states at both $y=1$ and $y=L_y$ surfaces. We have then investigated the FAs with further confinement in another direction which leads to a NW geometry. In WSM NW setup, due to the finite size effects, both the bulk and surface states are gapped out. Interestingly, the surface state gap is still smaller than the bulk confinement gap, thus allowing one to probe only surface states and explore various transport signatures mediated due to only FAs. We also analyse the localization properties of FAs in the NW geometry and find the spin textures similar to that 
in WSM slab. To obtain the conductance, we have extracted the matrix elements within the scattering matrix formalism using the python package KWANT~\cite{Groth2014}. Specifically, we have computed the two-terminal conductance of the WSM NW using the Landauer formula and observed the conductance quantization in units of $2e^2/h$. To capture the signatures of FAs in AR, we have constructed a WSM/WSC NW hybrid junction and found the conductance using the BTK formula~\cite{Blonder1982,Das2016,Dumitrescu2015,Rosdahl2018}. 
We show that the signatures of the FAs can be separated out via the AR process too. Note that, the conductance in this hybrid setup is not quantized. Finally, we have investigated the stability of the conductance against random onsite disorder potential in both WSM NW and WSM/WSC NW hybrid junction and find our results to be robust against disorder strength upto a critical value 
$V_{\rm dis}^{\rm c}$.

Here we convey few comments as far as the experimental feasibility of our transport setups are concerned. Earlier theoretical works to capture the signature of FAs are mainly based on TRS-broken WSM~\cite{Baum2015,Kaladzhyan2019,Mukherjee2019,Martino2021}. In reality, TRS-broken WSM needs application of large magnetic field~\cite{Xiong2015}, whereas experimentally observed WSM phases are mostly ISB {\it e.g.} TaAs, TaP, NbAs, NbP etc.~\cite{Lv2015a,Lv2015b,Lv2015c,Lu2015,Xu2015a,Xu2015b,Xu2015c,Soluyanov2015,Moll2016,Wang2016,Xu_2015,Ojanen2013,Sun2015,Feng2016,Xu2016,Potter2014}. Specifically, there exists several works on Dirac semimetal, $\rm{Cd_3As_2}$, NW where WSM phase 
can be achieved by applying a strong external magnetic field ($\sim 1$ Tesla) with diameter of the NW 
$\sim 20-100$ nm~\cite{Wang2018,Lin2017,Li2015,Wang2016b}. Application of large external magnetic field can generate Landau levels and the phenomenon of chiral anomaly may affect the results. 
Hence, our work based on ISB WSM NW is out of such scope and carries potential from the practical point of view. Note that, the bulk gap in NW due to quantum confinement is observed to be $\sim 10\,$ meV with gap size being $\sim \hbar v_f \sqrt{\pi/S}$ ($v_f,S$ being the Fermi velocity  and cross section of the NW respectively). For a typical value of $v_f \sim 10^5 \, m/s$ in $\rm{Cd_3As_2}$ NW and $W\sim 20 \AA$ (for lattice constant $\sim 1\AA$), gap size comes out to be $\sim 100~{\rm{meV}}$ which closely resembles the bulk confinement gap, $E_g^{\rm Bulk}=0.05t \sim 50~{\rm{meV}}$ (assuming $t\sim$ eV), obtained in our numerical calculation of band structure. Therefore, the realization of our theoretical results regarding FA mediated transport is subjected to the chemical potential lying within the bulk confinement gap, i.e. $\mu \lesssim 50-100~{\rm{meV}}$. Finally, the realization of a WSC can be achieved by introducing a common $s$-wave superconductor like 
Al or Nb in close proximity to the WSM NW~\cite{Bachmann2017}. Thus, our proposal serves as a possible potential experimental testbed, offering experimentalists the opportunity and challenge
to validate our findings.


\subsection*{Acknowledgements}

A.\,P.~acknowledges Arijit Kundu, Sourin Das, Arnob Kumar Ghosh, and Pritam Chatterjee for stimulating discussions. A.\,P.~and A.\,S.~acknowledge SAMKHYA: High-Performance Computing facility provided by Institute of Physics, Bhubaneswar, for numerical computations. P.\,D.~acknowledges Department of Space, Government of India for all supports at Physical Research Laboratory (PRL) and Department of Science and Technology (DST), India (through SERB Start-up Research Grant (File no.\,SRG/2022/001121)) for the financial support.
\bibliography{bibfile}{}
\end{document}